\documentclass{PoS}
\usepackage{subfigure}

\newcommand{\bea}{\begin{eqnarray}}
\newcommand{\eea}{\end{eqnarray}} 
\definecolor{orange}{rgb}{1,0.5,0}
\newcommand{\beq}{\begin{equation}}
\newcommand{\eeq}{\end{equation}}

\newcommand{\tK}{\widetilde{K}}

\newcommand{\Tr}{\rm{Tr}}
\renewcommand{\a}{\alpha}

\newcommand{\tr}{{\rm Tr}}

\renewcommand{\b}{\beta}

\newcommand{\bx}{\mathbf{x}}
\newcommand{\by}{\mathbf{y}}
\newcommand{\bk}{\mathbf{k}}

\newcommand{\vx}{\bx}
\newcommand{\vy}{\by}

\newcommand{\vk}{\bk}

\newcommand{\m}{\mu}

\renewcommand{\k}{\kappa}

\newcommand{\D}{\Delta}

\renewcommand{\th}{\theta}

\newcommand{\oh}{\frac{1}{2}}

\newcommand{\dg}{\dagger}
\newcommand{\non}{\nonumber}

\newcommand{\rf}[1]{(\ref{#1})}

\newcommand{\pa}{\partial}

\title{Effective Polyakov line actions, and their solutions at finite chemical potential}

\ShortTitle{Effective Polyakov line actions}

\author{\speaker{Jeff Greensite}\thanks{This research is supported in part by the
U.S.\ Department of Energy under Grant No.\ DE-FG03-92ER40711.} \\
        Physics and Astronomy Department \\
        San Francisco State University \\
        San Francisco, CA 94132 USA \\
        E-mail: \email{greensit@stars.sfsu.edu}}

\abstract{I outline recent progress in the relative weights approach to deriving effective Polyakov line
actions from an underlying lattice gauge theory, and compare mean field and complex Langevin methods
for solving such theories at finite chemical potential.}

\FullConference{The 32nd International Symposium on Lattice Field Theory,\\
		23-28 June, 2014\\
		Columbia University New York, NY}

\begin{document}

\section{Introduction}

    Any attempt to simulate QCD at finite chemical potential $\m$ must somehow deal with the sign problem, i.e.\ the fact that at $\m>0$ the fermion determinant is complex, and straightforward importance sampling is impossible.  Our approach to this problem is to first map the gauge theory into a theory with many fewer degrees of freedom, namely, a Polyakov line or ``SU(3) spin'' model, by a method we refer to as ``relative weights'' \cite{Greensite:2014isa}.  We will then deal with the sign problem in two different ways:  first using the complex Langevin equation, following the method of \cite{Aarts:2011zn}, and also by a mean field approach, as discussed in \cite{Greensite:2012xv}.  We will find that these methods sometimes agree perfectly, and sometimes not.  I will discuss who is  right $-$ or who is wrong $-$ in the latter case.
    
\section{The Relative Weights Method}

       Start with lattice gauge theory and integrate out all d.o.f.\ subject to the constraint that the Polyakov
line holonomies are held fixed.  This defines the Polyakov line action (PLA) $S_P$.  In temporal gauge 
\beq
e^{S_P[U_{\bx}]}=    \int  DU_0(\bx,0) DU_k  D\phi ~ \left\{\prod_\bx \delta [U_{\bx}-U_0(\bx,0)]  \right\}
 e^{S_L} 
\label{SP}
\eeq
where $\phi$ denotes any matter fields, bosonic or fermionic, in the lattice action $S_L$.  We will avoid dynamical fermion simulations for now, and work instead with an SU(3) gauge-Higgs model with a fixed modulus ($\Omega \Omega^\dg=1$) Higgs field
\beq
   S_L = {\beta \over 3} \sum_{p} \rm{ReTr}[U(p)] 
 +  {\kappa \over 3} \sum_{x}\sum_{\m=1}^4 \rm{Re}\Bigl[\Omega^\dg(x) U_\m(x) \Omega(x+\hat{\m})\Bigr]
\eeq
If we can derive $S_P$ at $\m=0$, then (in principle) we also have $S_P$ at $\m>0$ by the following identity: 
\beq      
S_P^\m [U_\bx,U^\dg_\bx] =  S_P^{\m=0}\Bigl[e^{N_t \m} U_\bx,e^{-N_t \m}U^\dg_\bx \Bigr]
\label{mu}
\eeq
which is true to all orders in the strong coupling/hopping parameter expansion.  This identity will be supplemented by simulations with an imaginary chemical potential, as explained below.

   Let $S'_L$ be the lattice action in temporal gauge with $U_0(\vx,0)$ fixed to $U'_\vx$.  It is not so easy to compute $S_P[U']$ directly.  But the ratio (``relative weights'')
\beq e^{\D S_P} = {\exp[S_P[U'_\vx]] \over \exp[S_P[U''_\vx]]}  \eeq
is easily computed as an expectation value
\bea
\exp[\D S_P] &=& {\int  DU_k  D\phi ~  e^{S'_L} \over \int  DU_k  D\phi ~  e^{S''_L} }
\non \\
&=&{\int  DU_k  D\phi ~  \exp[S'_L-S''_L] e^{S''_L} \over \int  DU_k  D\phi ~  e^{S''_L} }
\non \\
&=& \Bigl\langle  \exp[S'_L-S''_L] \Bigr\rangle'' 
\eea
where $U''_\vx$ denotes a configuration slightly different from $U'$, and $\langle ... \rangle''$ means the VEV in the Boltzman weight $\propto e^{S''_L}$. Now suppose $U_\vx(\lambda)$ is some path through configuration space parametrized by 
$\lambda$, and suppose $U'_\vx$ and $U''_\vx$ differ by a small change in that parameter, i.e.  
\beq
U'_\bx = U_\bx(\lambda_0 + \oh \D \lambda)  ~~~,~~~
U''_\bx = U_\bx(\lambda_0 - \oh \D \lambda )
\eeq
Then the relative weights method gives us the derivative of the true effective action $S_P$ along the path:
\beq 
\left( {dS_P \over d\lambda} \right)_{\lambda=\lambda_0}  \approx  {\D S \over \D \lambda} 
\eeq
The question is: which derivatives will help us to determine $S_P$ itself?  

\begin{figure}[htb]
\subfigure[~$\beta=5.6$]  % caption for subfigure a
{   
 \label{b56}
 \includegraphics[scale=0.6]{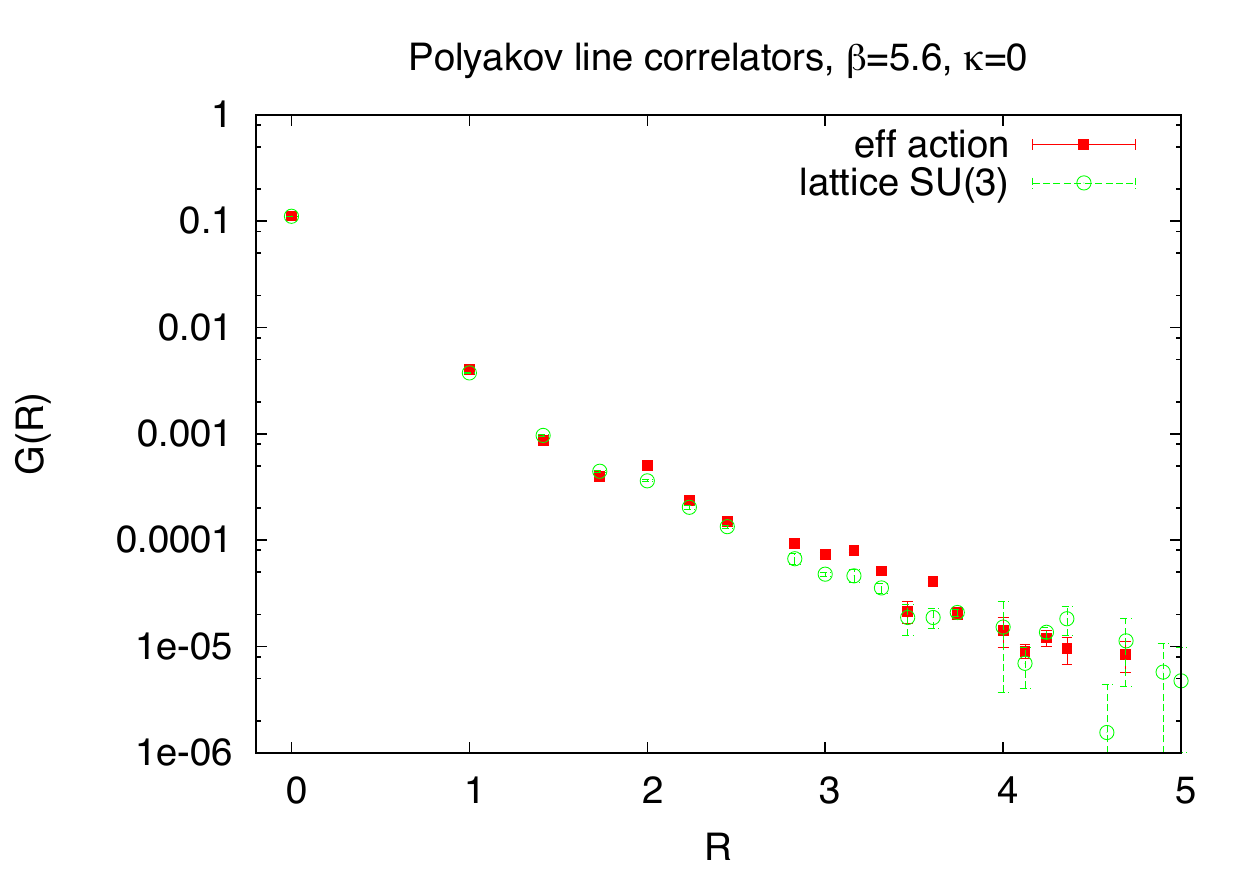}
}
\subfigure[~$\beta=5.7$]  % caption for subfigure a
{   
 \label{b57}
 \includegraphics[scale=0.6]{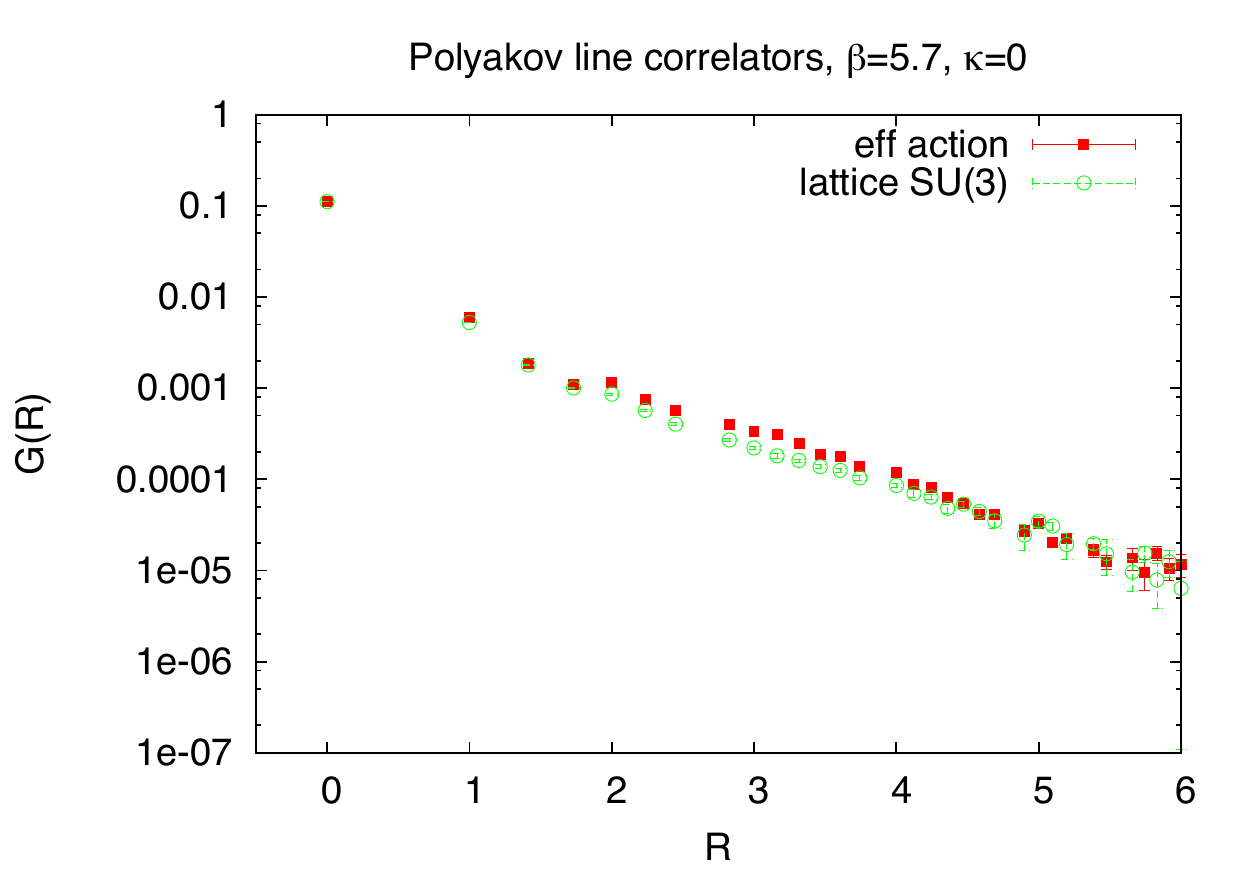}
}
\caption{The Polyakov line correlators for pure gauge theory at $\b=5.6$ and $\b=5.7$,  computed from numerical simulation of the effective PLA $S_P$, and from simulation of the underlying lattice SU(3) gauge theory.}
\label{pg}
\end{figure}

We compute derivatives of $S_P$ w.r.t.\ Fourier components $a_\bk$ of the Polyakov lines
\beq 
P_\bx \equiv {1\over N_c} {\Tr} U_\vx = \sum_\bk a_\bk e^{i\bk \cdot \bx} 
\eeq
For a pure gauge theory, the part of $S_P$ bilinear in $P_x$ is constrained to have the form
\beq
             S_P = \sum_{\vx \vy} P_\vx P^\dg_\vy K(\vx-\vy) 
\label{SP1}
\eeq
Then, going over to Fourier modes
\beq
             {1\over \alpha}{1\over L^3}\left( {\pa S_P \over \pa a^R_{\vk}}\right)_{a_\vk = \a} = 2 \tK(\vk) 
\eeq
Having obtained $\tK(\vk)$ in SU(3) gauge theory by the relative weights method just described, we Fourier transform to obtain $K(\vx-\vy)$, simulate the PLA \rf{SP1} by standard methods, and compute the Polyakov line correlator  
$G(R) = \langle P_\vx P^\dg_\vy \rangle$. This correlator can also be computed in the underlying lattice pure gauge theory,
with the results (on a $16^3 \time 6$ lattice) shown in Fig.\ \ref{pg}.

\begin{figure}[htb]
\subfigure[~$\kappa=3.6$]  % caption for subfigure a
{   
 \label{k36}
 \includegraphics[scale=0.37]{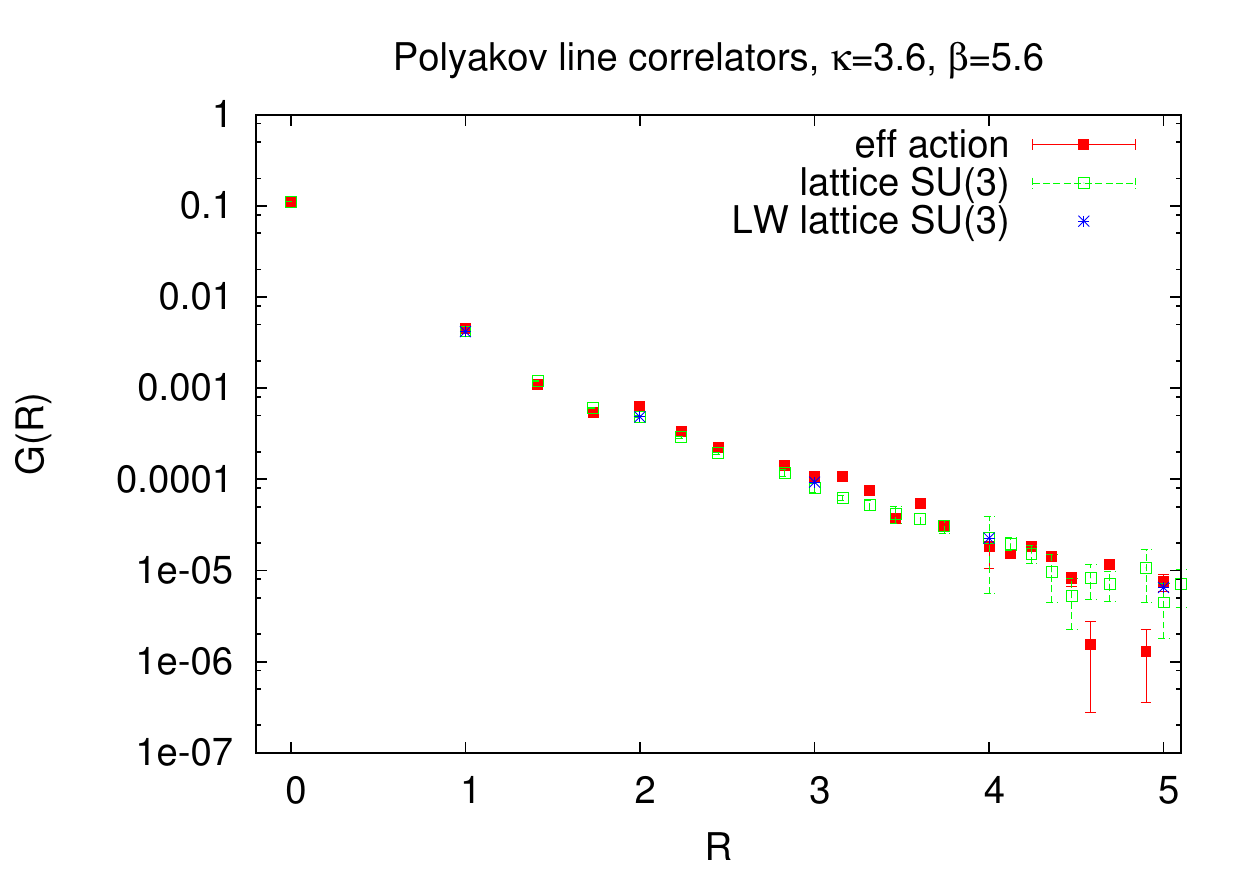}
}
\subfigure[~$\kappa=3.8$]  % caption for subfigure a
{   
 \label{k38}
 \includegraphics[scale=0.37]{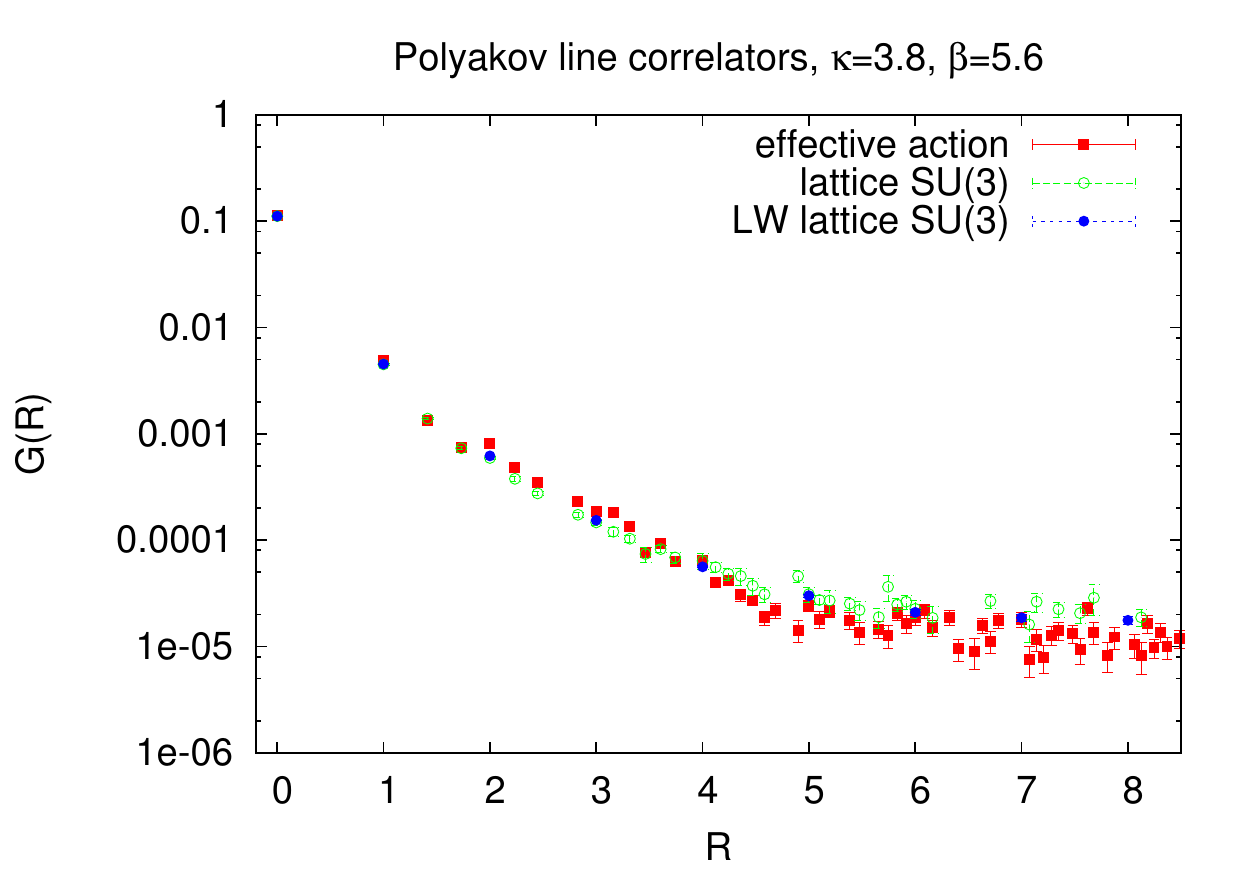}
}
\subfigure[~$\kappa=3.9$]  % caption for subfigure a
{   
 \label{k39}
 \includegraphics[scale=0.37]{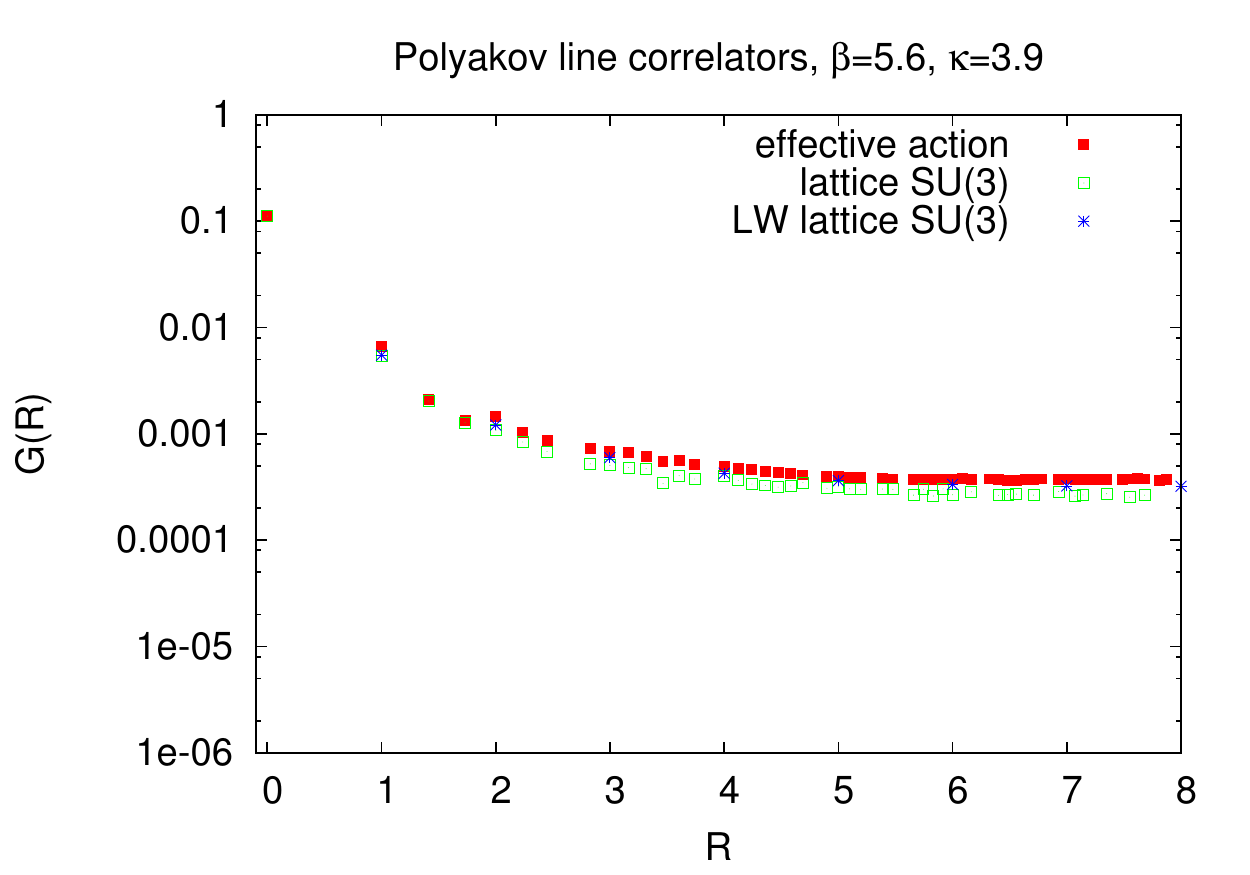}
}
\caption{The Polyakov line correlators for the gauge-Higgs theory at $\beta=5.6$ and $\kappa=3.6,3.7,3.8$,  computed from numerical simulation of the effective PLA $S_P$, and from simulation of the underlying lattice SU(3) gauge theory. In the latter case we show off-axis points computed by standard methods, together with on-axis points using L\"uscher-Weisz noise reduction.}
\label{gh}
\end{figure}

  We now consider the SU(3) gauge-Higgs action.  Including linear and bilinear center symmetry-breaking terms, it can be shown that at finite chemical potential
\bea
S_P &=& \sum_{\vx \vy} P_\vx P_\vy^\dg K(\vx-\vy) + 
   \sum_{\vx \vy} (P_\vx P_\vy Q(\vx-\vy,\mu) +  P^\dg_\vx P^\dg_\vy Q(\vx-\vy;-\m) )
\non \\
& & + \sum_\vx \Bigl\{(d_1 e^{\mu/T} - d_2 e^{-2\mu/T}) P_\vx + (d_1 e^{-\mu/T} - 
d_2 e^{2\mu/T}) P^\dg_\vx \Bigr\}
\label{ghmu}
\eea
To help determine center symmetry-breaking coefficients $d_1,d_2,Q(\vx-\vy;\mu)$ it is useful to compute $dS_P/da_\vk$ at imaginary chemical potential $\mu/T=i\theta$.  This resolves certain ambiguities in the application of \rf{mu}.  Details can be found in ref.\ \cite{Greensite:2014isa}.  In our exploratory work in \cite{Greensite:2014isa} we have neglected $Q(\vx-\vy;\mu)$ on the grounds that it is rather small;
however, the existence of quadratic symmetry-breaking terms that may be important at large $\m$ can be inferred (see below).  Having determined $K(\vx-\vy)$ and $d_1,d_2$ by the relative weights method, we can then compare, at $\m=0$, Polyakov line correlators computed by Monte Carlo simulation of the PLA and of the underlying gauge-Higgs theory.  The comparison is shown, at $\b=5.6$ and several values of $\k$, in Fig.\ \ref{gh}.

\section{Solutions at finite chemical potential}

    Our next step is to try to solve the effective PLA at non-zero $\m$.  We first ignore
bilinear symmetry-breaking terms, and solve \rf{ghmu} with $Q=0$.  A variant is to consider that the symmetry-breaking
terms proportional to $d_2$ most naturally arise from ``double-winding'' terms
\beq
 d_2 e^{2\m/T} \tr[U_\vx^2] + d_2 e^{-2\m/T} \tr[U_\vx^{\dg 2}]
\eeq
via the SU(3) identities
\beq
\tr[U_\vx^2] =\tr[U_\vx]^2 -  2\tr[U^\dg_\vx]  ~~~,~~~ \tr[U_\vx^{\dg 2}] =\tr[U^{\dg}_\vx]^2 -  2\tr[U_\vx] \ ,
\eeq
With that motivation we will also consider the action
\bea
S_P &=& {1\over 9}\sum_{xy} \tr[U_\vx] \tr[U_\vy^\dg] K(\vx-\vy) 
 + {1\over 3} \sum_x \Bigl\{d_1 e^{\m/T} \tr[U_\vx] +  
         d_1 e^{-\m/T}  \tr[U^\dg_\vx] \Bigr\} 
\non \\
& & + {1\over 6} \sum_x \Bigl\{d_2 e^{2\m/T} \tr[U_\vx^2] +  
         d_2 e^{-2\m/T}  \tr[U^{\dg 2}_\vx] \Bigr\} 
\label{dwind}
\eea
Lastly, we can consider the ``heavy-dense'' quark model in temporal gauge:
\beq
e^{S_L} = \prod_\vx \det\Bigl[1+h e^{\m/T} U_0(\vx,0)\Bigr]^p \det\Bigl[1+h e^{-\m/T} U^\dg(\vx,0) \Bigr]^p   e^{S_{plaq}}
\eeq
where $p=1$ for staggered fermions, $p=2N_f$ for Wilson fermions.  If we compute the Polyakov line action $S_P^{pg}$ for the pure gauge theory via relative weights, then
\beq
e^{S_P} = \prod_\vx \det\Bigl[1+h e^{\m/T} U_\vx \Bigr]^p \det\Bigl[1+h e^{-\m/T} U^\dg_\vx \Bigr]^p   e^{S_P^{pg}}
\eeq

\begin{figure}[htb]
\subfigure[~$\langle \tr(U)\rangle$]  % caption for subfigure a
{   
 \label{uhq}
 \includegraphics[scale=0.37]{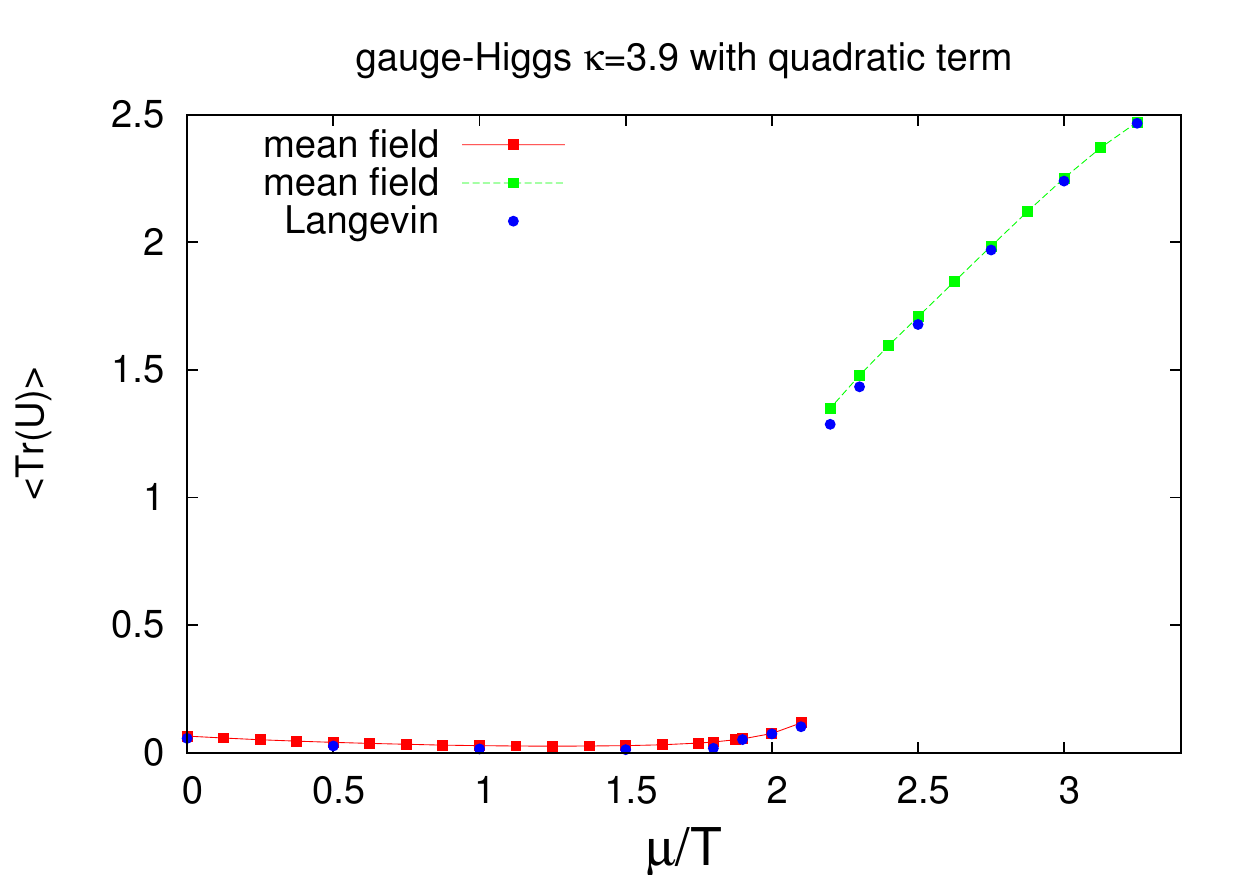}
}
\subfigure[~$\langle \tr(U^\dg)\rangle$]  % caption for subfigure a
{   
 \label{vhq}
 \includegraphics[scale=0.37]{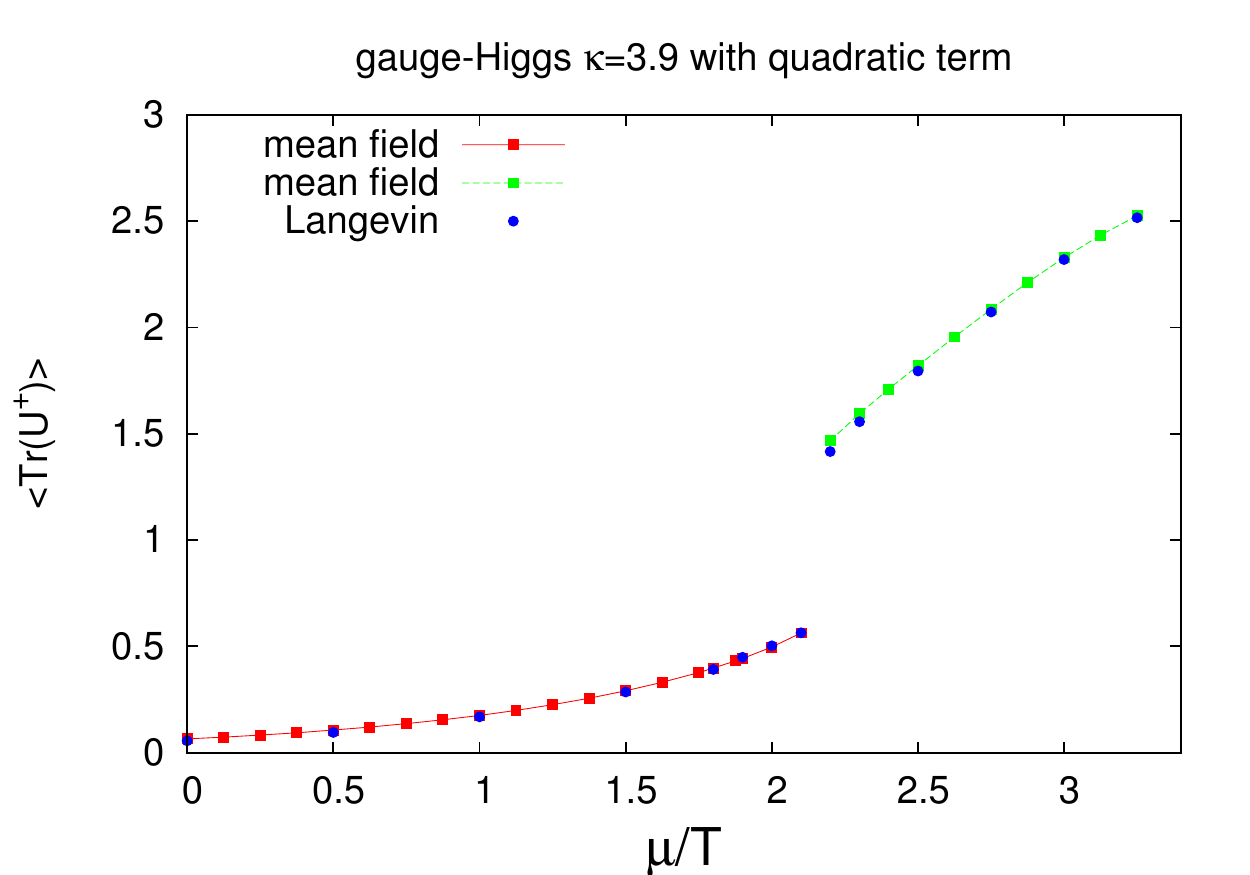}
}
\subfigure[~density]  % caption for subfigure a
{   
 \label{nhq}
 \includegraphics[scale=0.37]{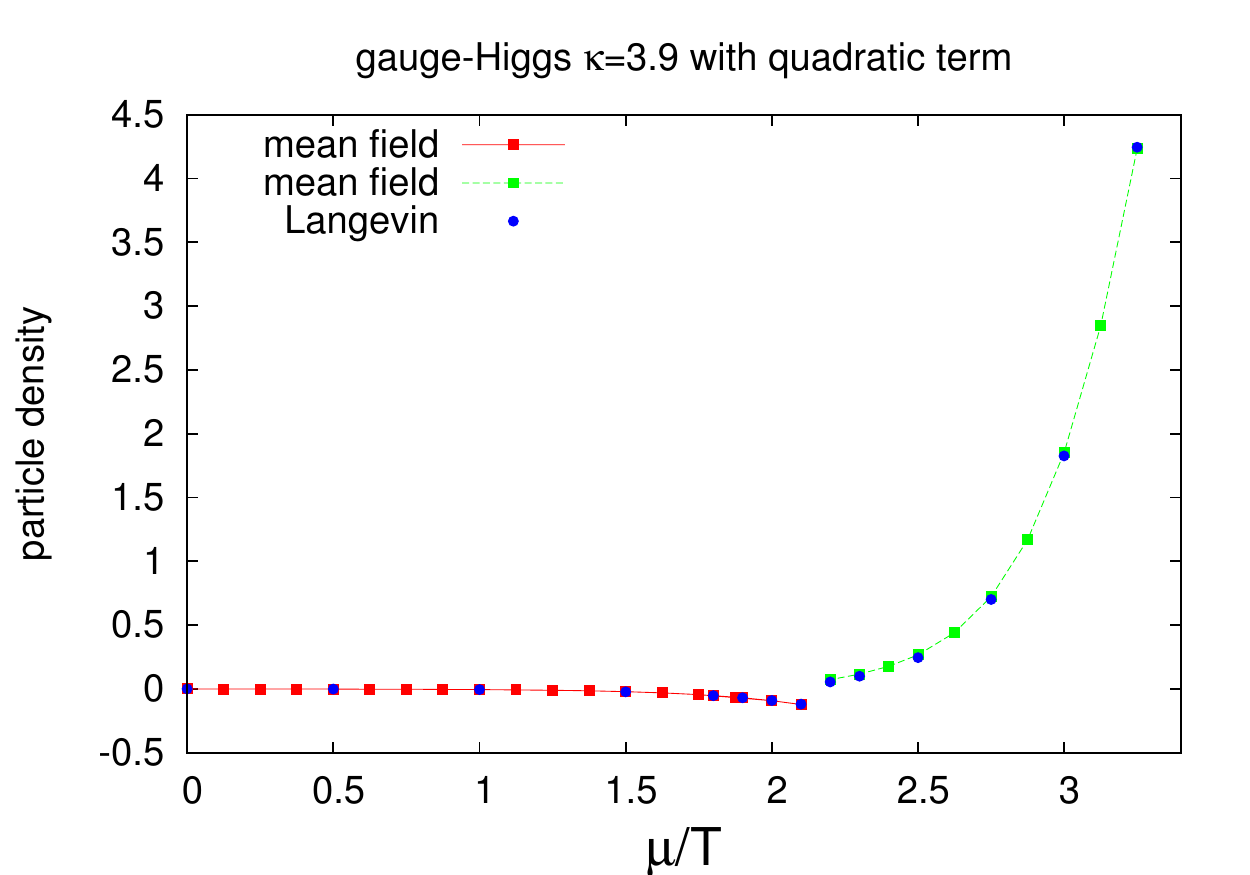}
}
\caption{Comparison of Polyakov lines $\langle \tr(U) \rangle, \langle \tr(U^\dg) \rangle$ and number density vs.\ $\m/T$, computed via  complex Langevin and mean field techniques, in gauge-Higgs theory at $\kappa=3.9$ for the action $S_P$ in eq.\ (3.3), which includes quadratic center symmetry-breaking terms.}
\label{dwindfig}
\end{figure}

   We solve these theories by complex Langevin, following the approach of Aarts and James \cite{Aarts:2011zn}, and also by a mean field method \cite{Greensite:2012xv}.  Let the three eigenvalues of a Polyakov line holonomy be $e^{i\th_i}$ with $\th_3=-(\th_1+\th_2)$,
and the logarithm of the Haar integration measure becomes part of the Lagrangian.
The angles $\th_{1,2}$ are complexified, and the complex Langevin equation is solved numerically.  However, as pointed out by M{\o}llgaard and Splittorff \cite{Mollgaard:2013qra}, the complex Langevin approach can lead to incorrect results if the evolution repeatedly crosses the branch cut of the logarithm.  To study this, we keep track of the argument of the logarithm of the Haar measure
\beq
\mbox{Arg} = \sin^2\left({\theta_1(\vx')-\theta_2(\vx') \over 2}\right)\sin^2\left({2\theta_1(\vx')+\theta_2(\vx') \over 2}\right)\sin^2\left({\theta_1(\vx')+2\theta_2(\vx') \over 2}\right) 
\label{arg}
\eeq
at an an arbitrary lattice site $\vx'$.  In Fig.\ \ref{dwindfig} I show the results for the Polyakov lines and the number density for the action \rf{dwind}, derived from complex Langevin and mean field at $\beta=5.6, \k=3.9$.  Note the phase transition. It is hard to even detect a difference between the two methods.  One also finds that the argument of the logarithm \rf{arg} almost never crosses the branch cut.\footnote{However, at the larger $\m$ values one also finds that complex Langevin evolution has more than one solution, depending on initial conditions.  It is necessary to choose the solution which has
a probability distribution which is bounded by an exponential dropoff in the space of complexified angles.}

\setcounter{subfigure}{0}

\begin{figure}[htb]
\subfigure[~$\langle \tr(U)\rangle$]  % caption for subfigure a
{   
 \label{uhq}
 \includegraphics[scale=0.37]{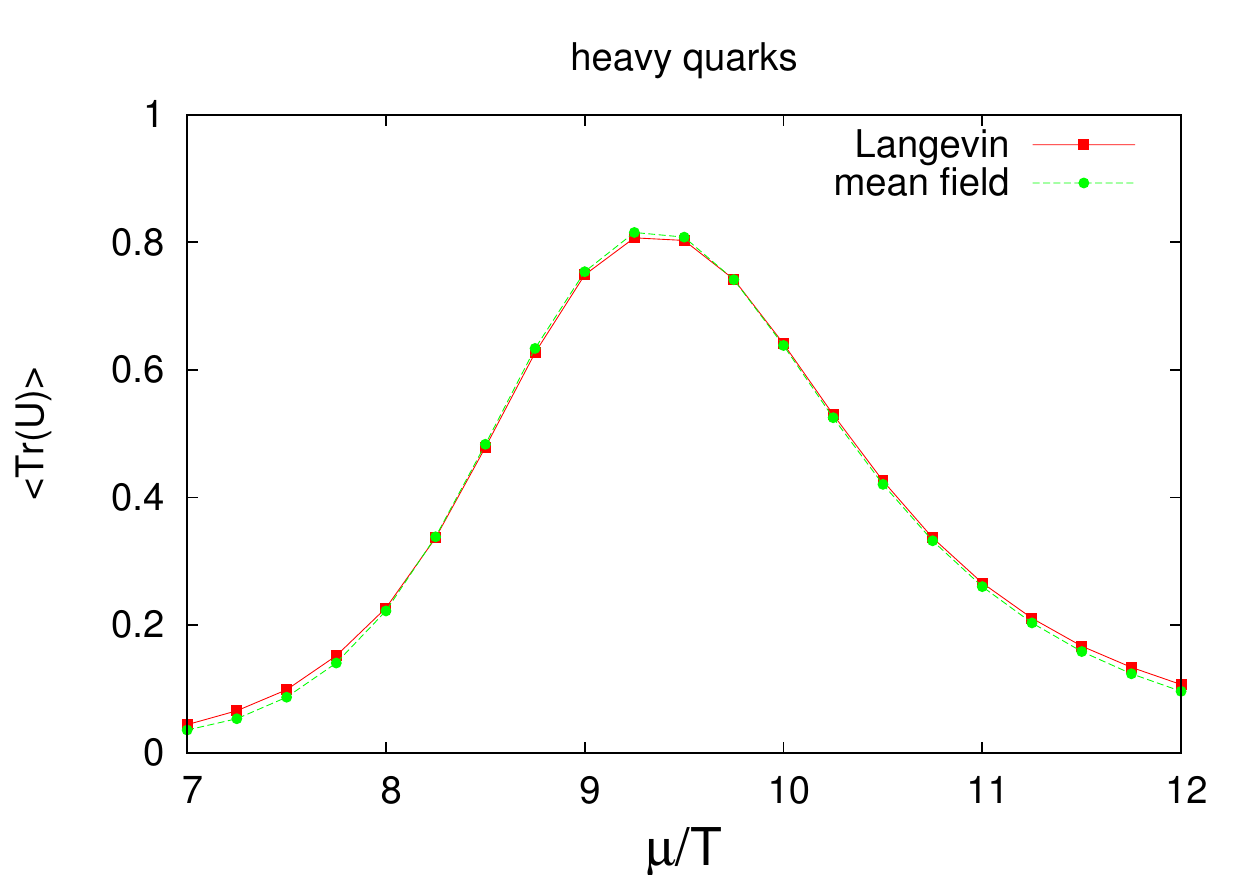}
}
\subfigure[~$\langle \tr(U^\dg)\rangle$]  % caption for subfigure a
{   
 \label{vhq}
 \includegraphics[scale=0.37]{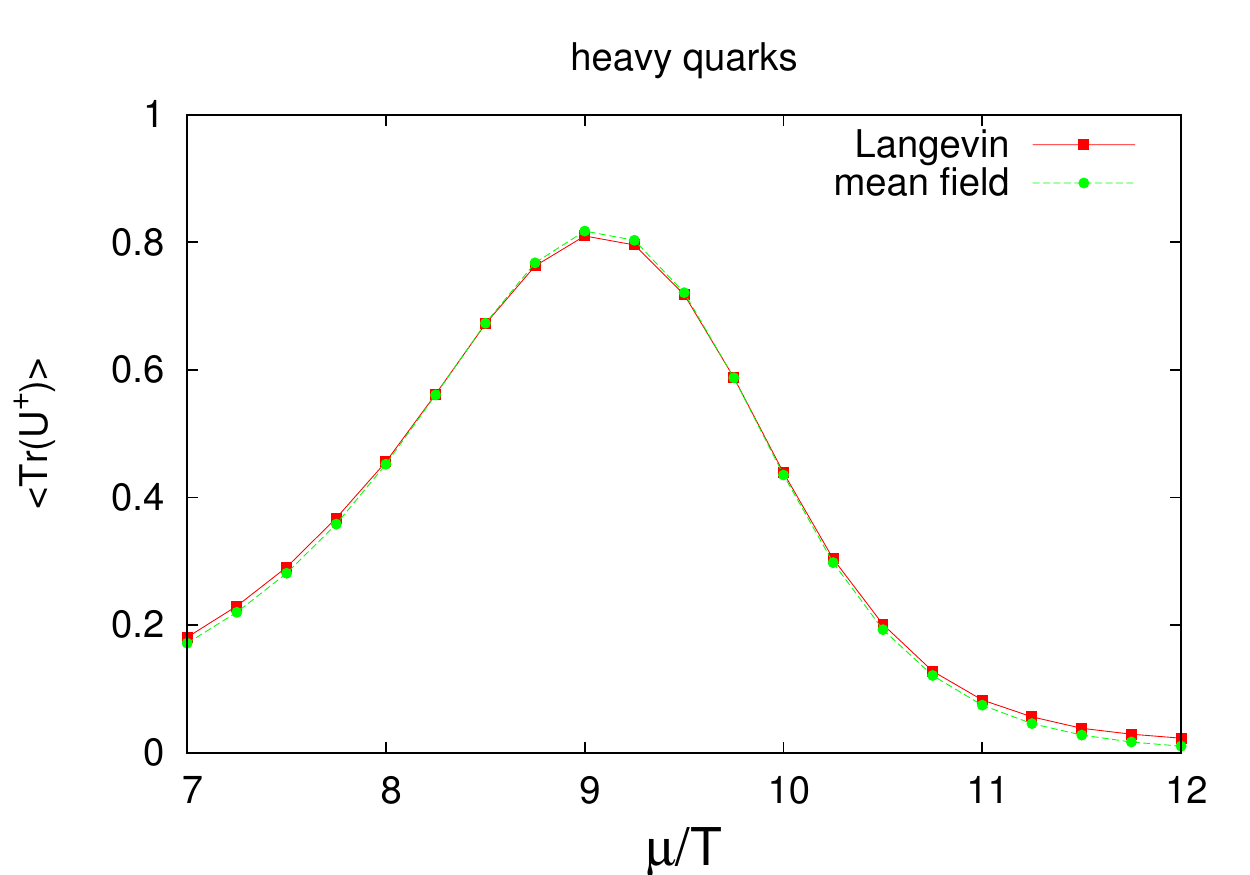}
}
\subfigure[~density]  % caption for subfigure a
{   
 \label{nhq}
 \includegraphics[scale=0.37]{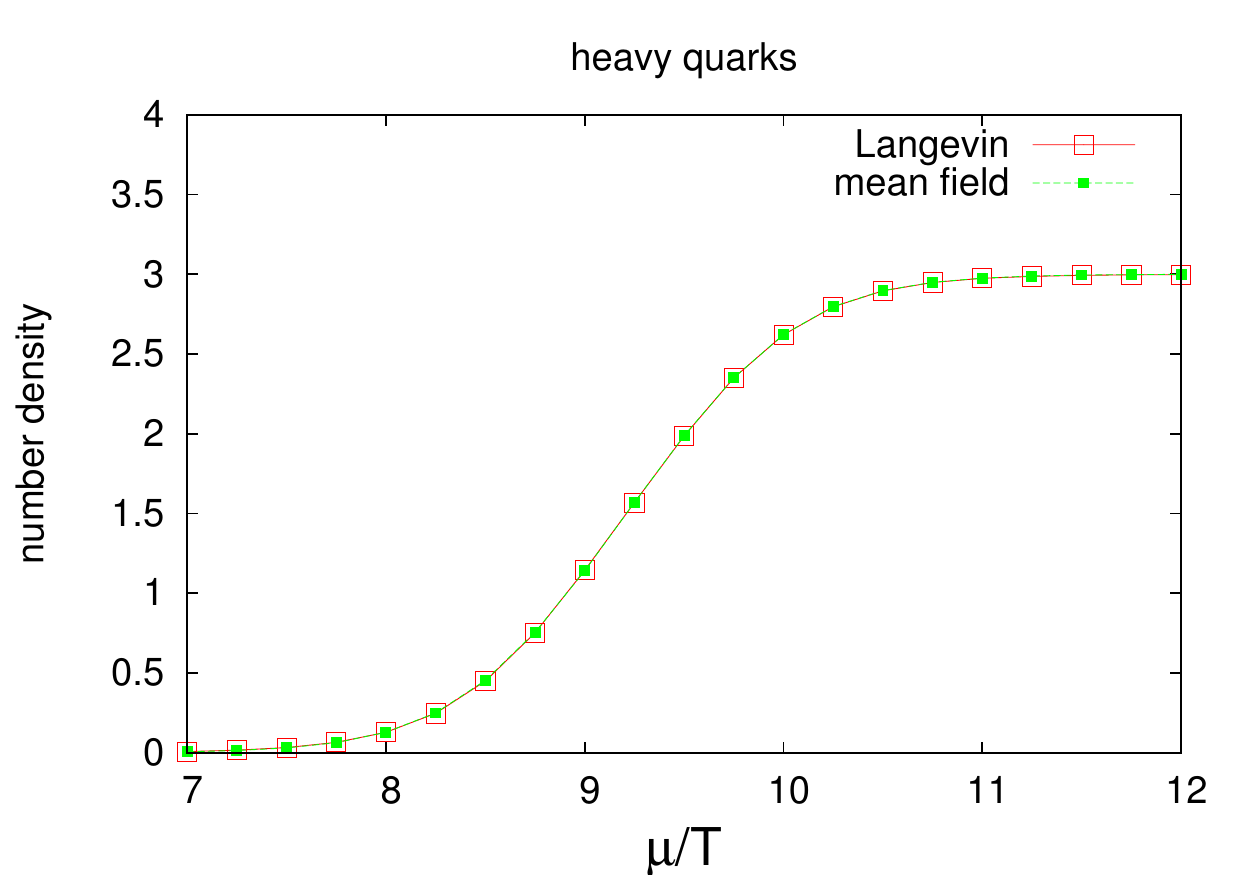}
}
\caption{Comparison of Polyakov lines $\langle \tr(U) \rangle, \langle \tr(U^\dg) \rangle$ and number density vs.\ $\mu/T$, computed via  complex Langevin and mean field techniques in the heavy-dense quark model. Note the saturation at high $\m/T$
at density=3.}
\label{quark}
\end{figure}

   Likewise, for the heavy-dense quark model, there is also near-perfect agreement between the two methods, as seen in Fig.\ \ref{quark}, and one also finds no branch-cut crossing problem. (For other approaches to this model, see e.g.\  \cite{Sexty:2013ica}, \cite{Langelage:2014vpa}.)
\setcounter{subfigure}{0}

\begin{figure}[htb]
\subfigure[~$\langle \tr(U) \rangle$]  % caption for subfigure a
{   
 \label{u39}
 \includegraphics[scale=0.37]{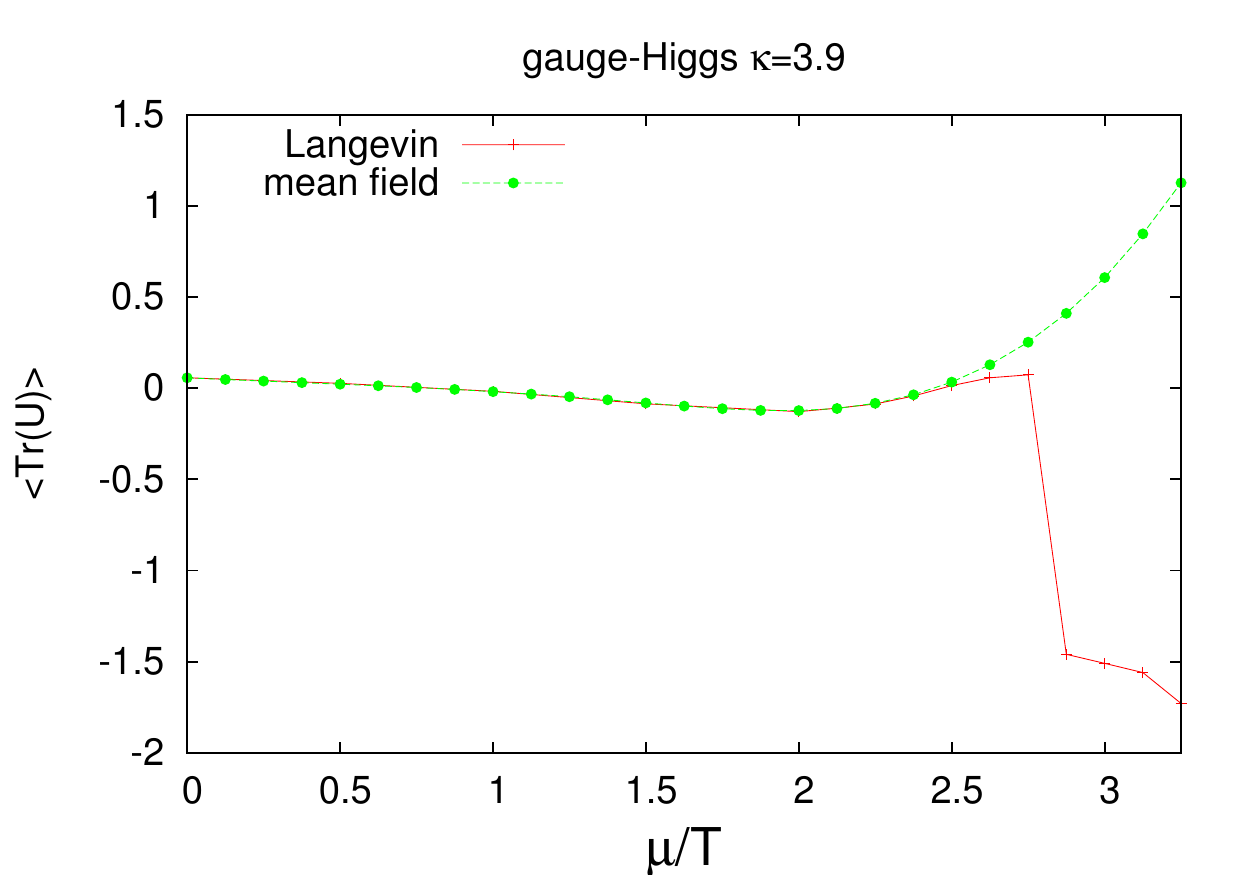}
}
\subfigure[~$\langle \tr(U^\dg) \rangle$]  % caption for subfigure a
{   
 \label{v39}
 \includegraphics[scale=0.37]{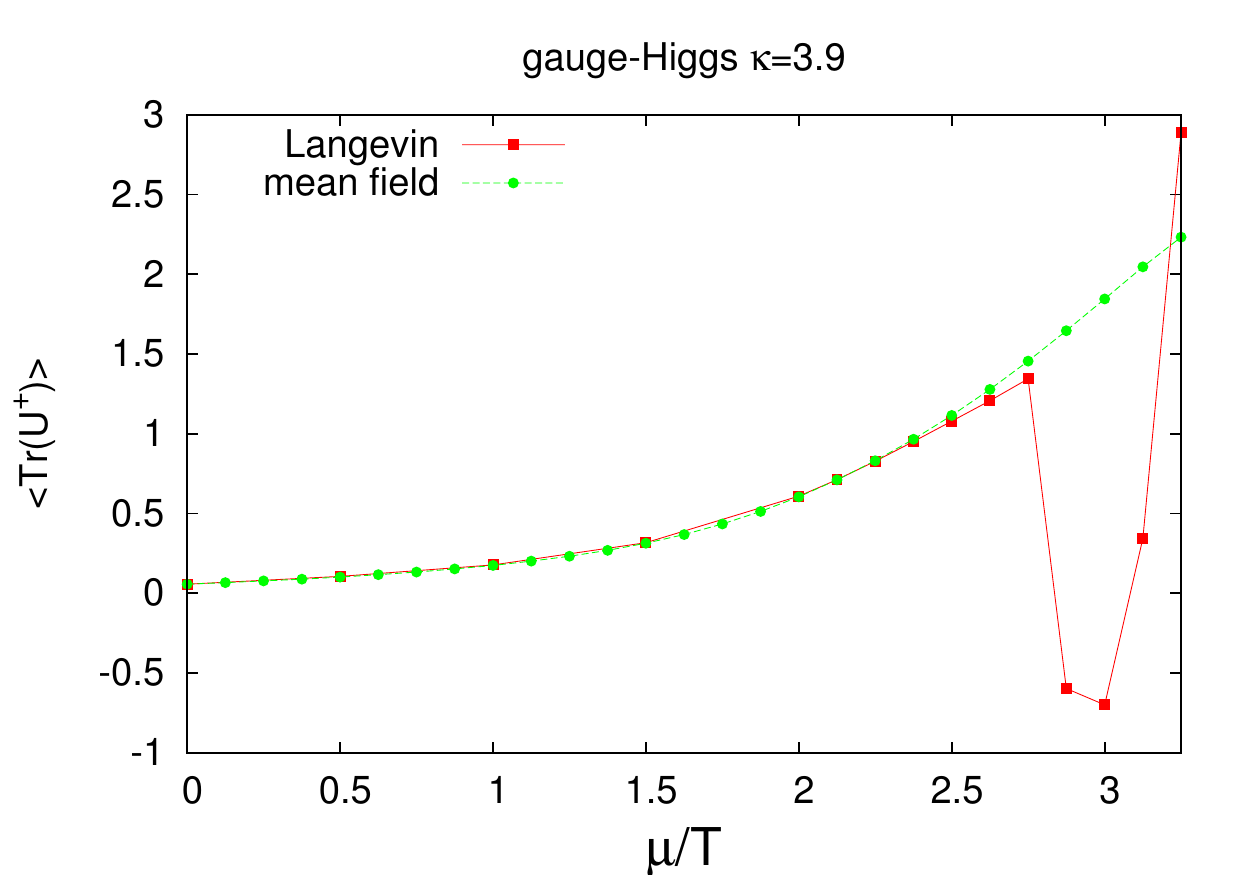}
}
\subfigure[~density]  % caption for subfigure a
{   
 \label{n39}
 \includegraphics[scale=0.37]{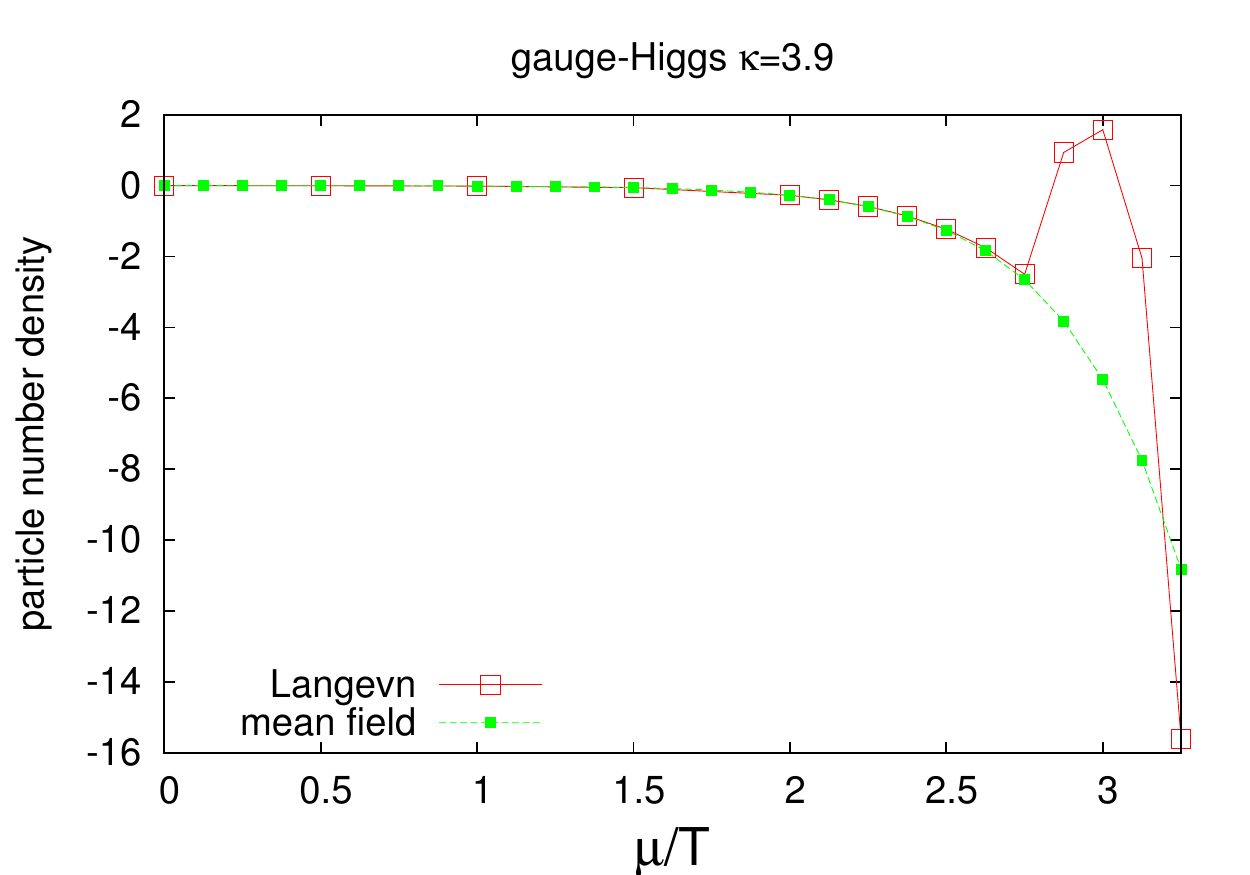}
}
\caption{Comparison of Polyakov lines $\langle \tr(U) \rangle, \langle \tr(U^\dg) \rangle$ and number density vs.\ $\m/T$, computed via  complex Langevin and mean field techniques, in gauge-Higgs theory at $\kappa=3.9$ for the action $S_P$ in eq.\ (2.11), where quadratic center symmetry-breaking terms are neglected.}
\label{ghfig}
\end{figure}

\begin{figure}[htb]
\subfigure[~$\m/T=1.5$]  % caption for subfigure a
{   
 \label{}
 \includegraphics[scale=0.37]{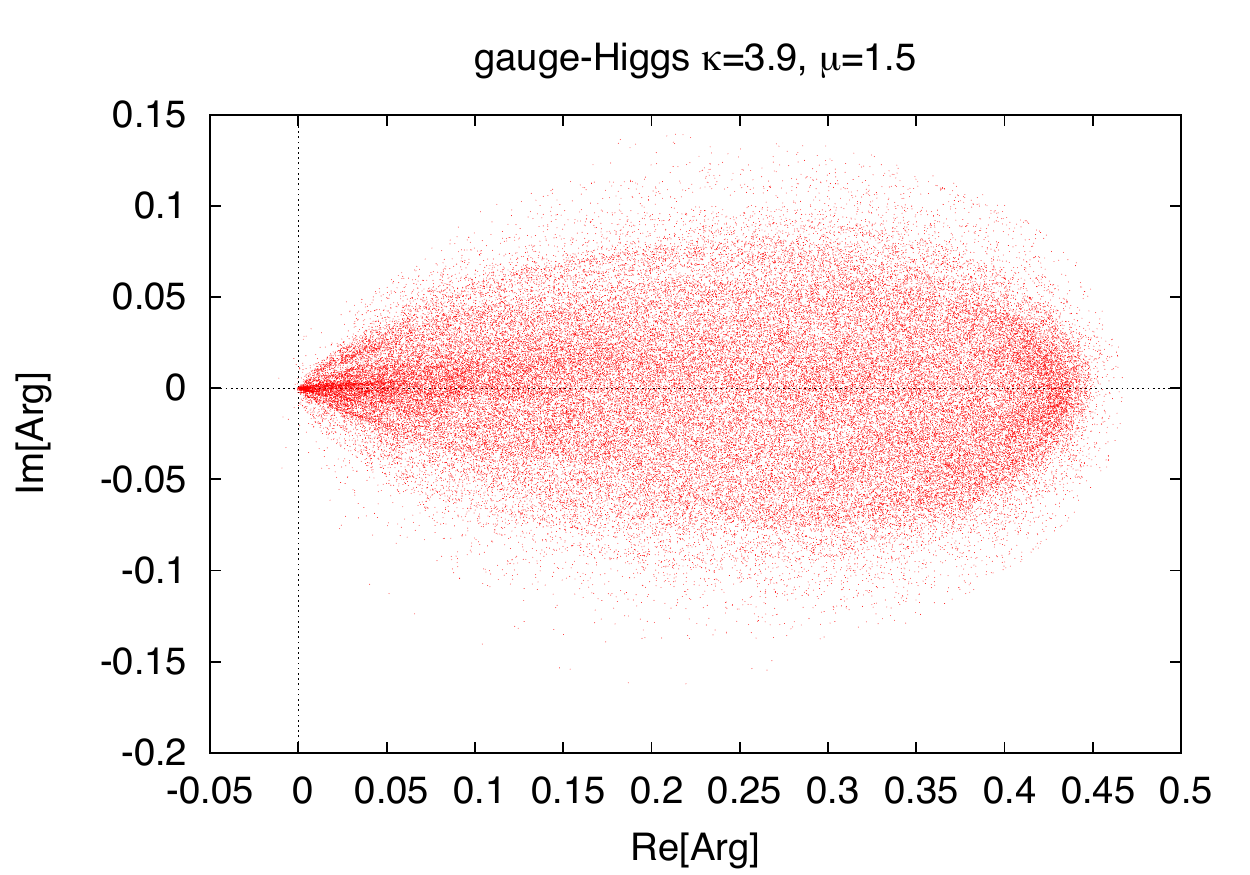}
}
\subfigure[~$\m/T=2.0$]  % caption for subfigure a
{   
 \label{}
 \includegraphics[scale=0.37]{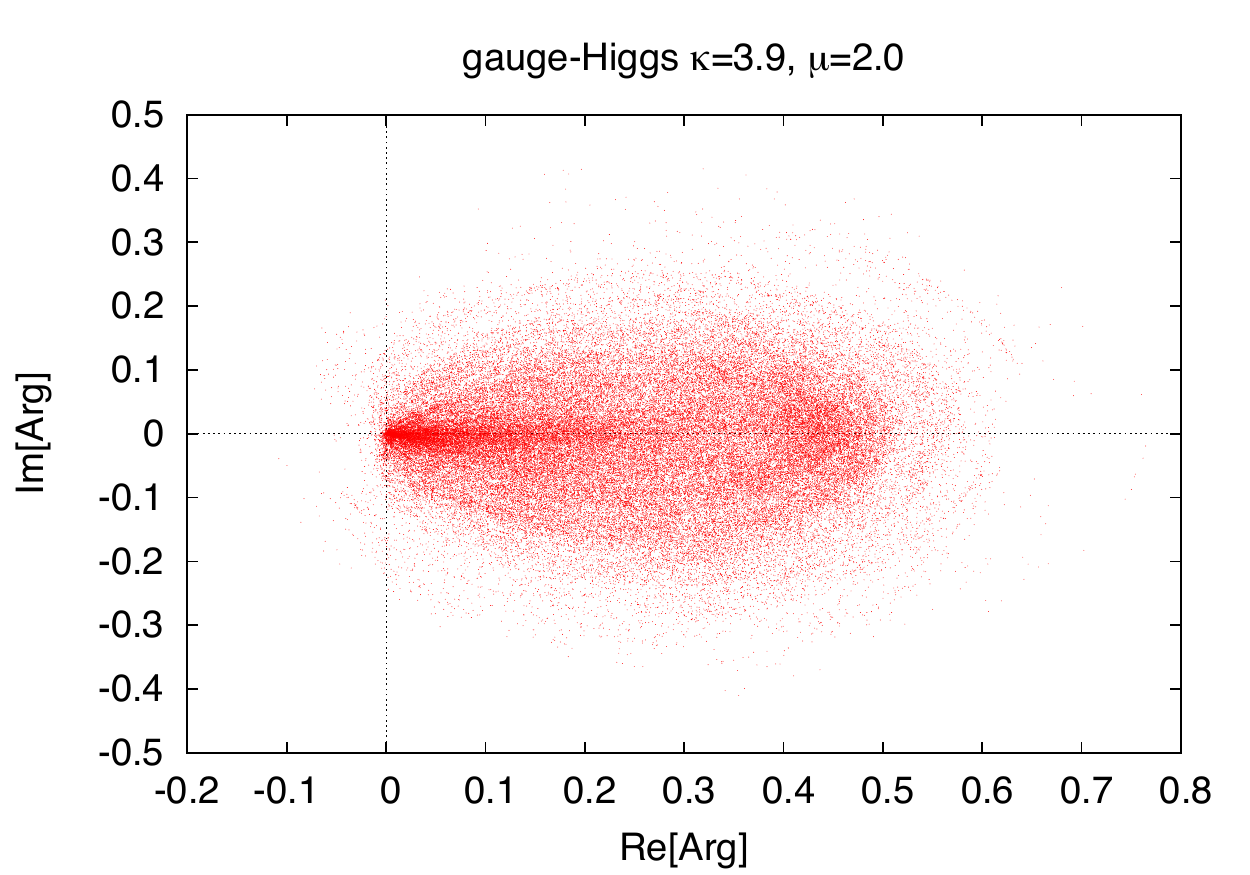}
}
\subfigure[~$\m/T=2.75$]  % caption for subfigure a
{   
 \label{}
 \includegraphics[scale=0.37]{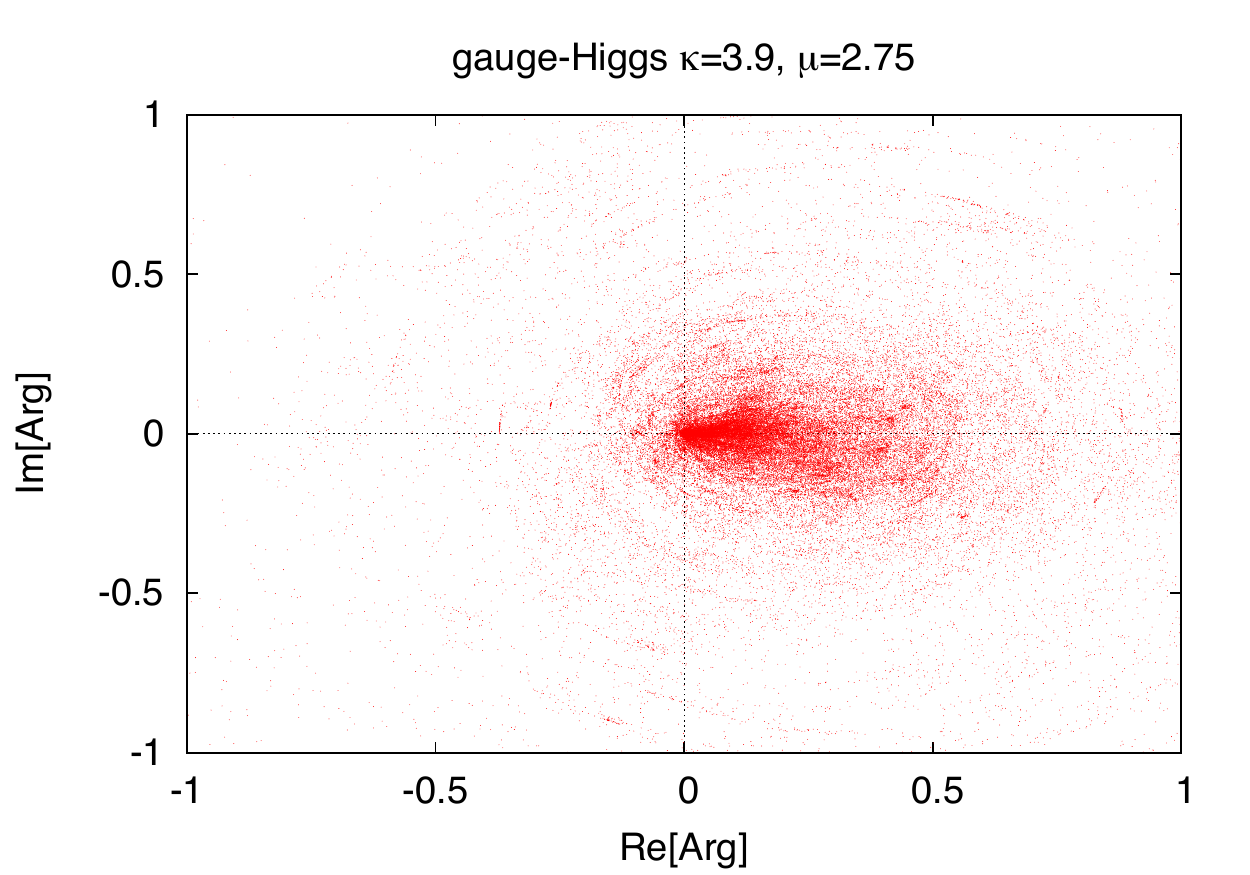}
}
\subfigure[~$\m/T=3.25$]  % caption for subfigure a
{   
 \label{}
 \includegraphics[scale=0.37]{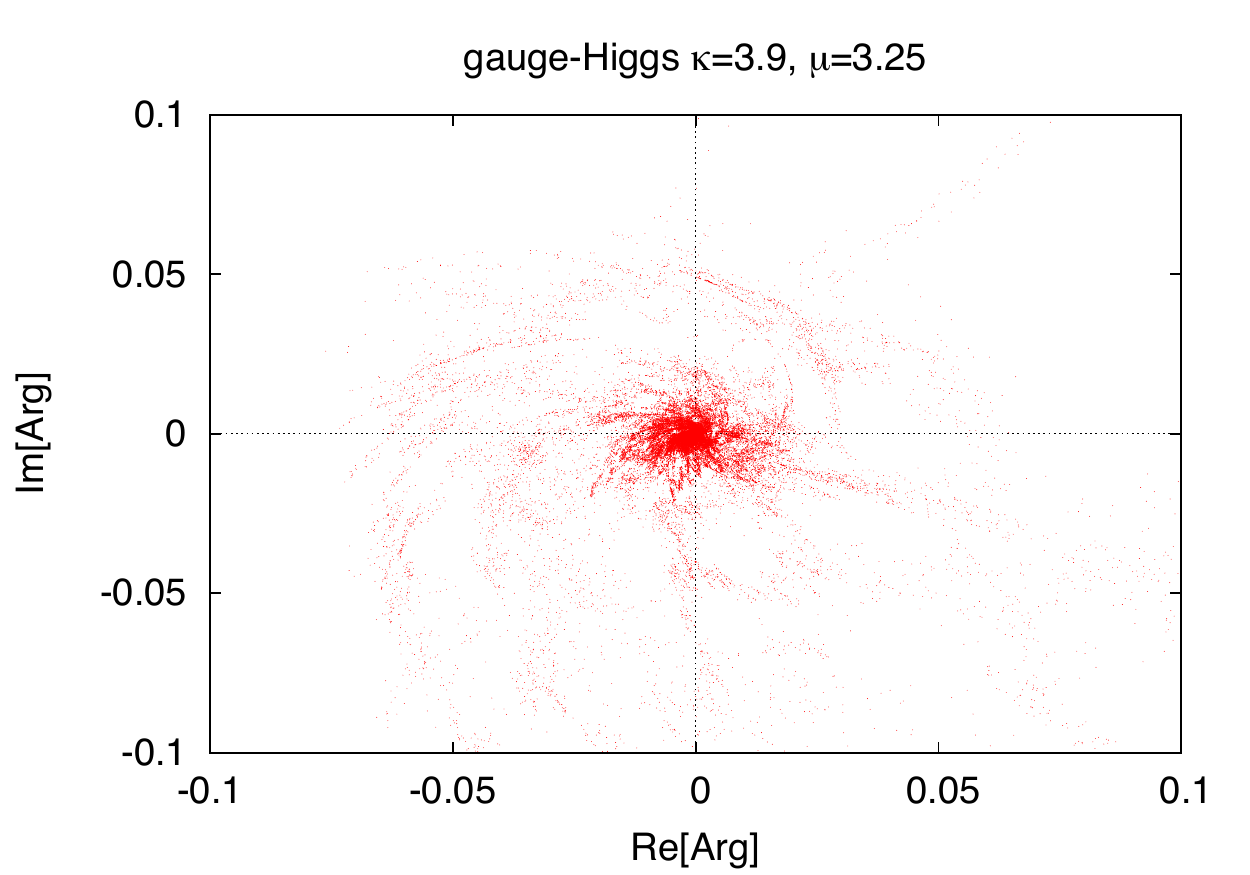}
}
\caption{Argument of the logarithm for gauge-Higgs theory at $\b=5.6,~\k=3.9$, and chemical potentials
$1.5 \le \m/T \le 3.25$ (subfigures a-d), evaluated at each Langevin time step.  The presence of many points near the negative real axis is very plain at $\b \ge 2.75$, signaling the presence of a branch cut problem.}
\label{det}
\end{figure}

   When we consider the action \rf{ghmu} at $\b=5.6,\k=3.9$, the Langevin and mean-field results diverge
at approximate $\m=2.75$, as seen in Fig.\ \ref{ghfig}.
\setcounter{subfigure}{0}
However, where the results differ, it turns out that complex Langevin evolution has a branch-cut crossing problem
of the type pointed out by M{\o}llgaard and Splittorff \cite{Mollgaard:2013qra}, as seen
in a plot (Fig.\ \ref{det}) of \rf{arg} at various values of $\m$.
\setcounter{subfigure}{0}
 
    It is natural to ask why mean field works so well.  This is probably due to the fact that  {\it many} spins, not just nearest neighbors, are coupled to a given spin, through  the non-local kernel $K(\vx-\vy)$.  For this reason the basic idea behind mean field theory, i.e.\ that each spin is effectively coupled to the average spin on the lattice, may be a very good approximation to the true situation.
 
 \section{Conclusions}
 
    To summarize:  We have developed a method for determining the effective Polyakov line action, at both zero and
finite chemical potential $\mu$.  At $\m=0$ there is excellent agreement for the Polyakov line correlators
computed in the effective theory and underlying lattice gauge theory.  At $\m > 0$ we can solve the effective theory by either mean field or complex Langevin methods.   Where the two methods agree, they agree almost perfectly.  Where they disagree, complex Langevin has a M{\o}llgaard-Splittorff branch cut crossing problem \cite{Mollgaard:2013qra}, which demonstrates that a problem of this sort can arise in a field theory whose action includes a logarithmic part.

   A possible way around the branch cut difficulty for Polyakov line models is to complexify the SU(3) elements $U_\vx, U^\dg_\vx$, rather than the angles $\th_a(\vx)$, a strategy which is used for lattice gauge theory (see \cite{Aarts:2013naa} for a review) and which was already mentioned in \cite{Aarts:2011zn}.  In that case the exponentiation of the measure factor is avoided, and there is no branch cut problem.  It is interesting nevertheless to see that the branch cut problem does appear in a non-trivial field theory with a 
logarithmic determinant in the action.  In real QCD, it is impossible to avoid the logarithm of the fermionic determinant in the action.  Whether the branch cut problem will
be an issue for Langevin simulations of real QCD  with light fermions is as yet unknown.

  The next step in this project is to apply the relative weights technique to a gauge theory coupled to dynamical fermions, and again the goal is to extract the effective Polyakov action via the relative weights method, as solve it at finite $\m$ via mean field theory.  We hope to report on the results at a later time.

   For a more complete exposition of the work described in these proceedings, see
\cite{Greensite:2014isa} and \cite{Greensite:2014cxa}.

\bibliography{pline}  

\begin{thebibliography}{1}

\bibitem{Greensite:2014isa}
J.~Greensite and K.~Langfeld,
\newblock Phys.Rev. {\bf D90}, 014507 (2014), arXiv:1403.5844.
%%CITATION = ARXIV:1403.5844;%%

\bibitem{Aarts:2011zn}
G.~Aarts and F.~A. James,
\newblock JHEP {\bf 1201}, 118 (2012), arXiv:1112.4655.
%%CITATION = ARXIV:1112.4655;%%

\bibitem{Greensite:2012xv}
J.~Greensite and K.~Splittorff,
\newblock Phys.Rev. {\bf D86}, 074501 (2012), arXiv:1206.1159.
%%CITATION = ARXIV:1206.1159;%%

\bibitem{Mollgaard:2013qra}
A.~Mollgaard and K.~Splittorff,
\newblock Phys.Rev. {\bf D88}, 116007 (2013), arXiv:1309.4335.
%%CITATION = ARXIV:1309.4335;%%

\bibitem{Sexty:2013ica}
D.~Sexty,
\newblock Phys.Lett. {\bf B729}, 108 (2014), arXiv:1307.7748.
%%CITATION = ARXIV:1307.7748;%%

\bibitem{Langelage:2014vpa}
J.~Langelage, M.~Neuman, and O.~Philipsen,
\newblock JHEP {\bf 1409}, 131 (2014), arXiv:1403.4162.
%%CITATION = ARXIV:1403.4162;%%

\bibitem{Aarts:2013naa}
G.~Aarts,
\newblock (2013), arXiv:1312.0968.
%%CITATION = ARXIV:1312.0968;%%

\bibitem{Greensite:2014cxa}
J.~Greensite,
\newblock (2014), arXiv:1406.4558.
%%CITATION = ARXIV:1406.4558;%%

\end{thebibliography}
 
\end{document}